\documentclass{article}

\def\mytitle#1{\setcounter{equation}{0}
\setcounter{footnote}{0}
\begin{flushleft}\Large\textbf{#1}\end{flushleft}
\vspace{0.25cm}}
\def\myname#1{\leftline{{\large #1}}\vspace{-0.13cm}}
\def\myplace#1#2{\small\begin{flushleft}\textit{#1}\\
\texttt{#2}\end{flushleft}}
\newenvironment{contribution}{\normalsize\noindent}{}

\usepackage{graphicx}% Include figure files
\begin{document}

\mytitle{Unified First Law and Some Comments}

\frenchspacing
\myname{ Subenoy Chakraborty*,~Ritabrata Biswas**,~Nairwita Mazumder$\dag$  }
\frenchspacing

\myplace{Department of Mathematics, Jadavpur University, Kolkata-32, India.}
{[*schakraborty@math.jdvu.ac.in~,**biswas.ritabrata@gmail.com~,~$\dag$nairwita15@gmail.com]}

\begin{abstract}
In this small article, unified first law has been analyzed and
some results have been deduced from it. The results have been
presented in the form of lemmas and some conclusions have been
drawn from them.

\end{abstract}

\begin{contribution}
\end{contribution}

The joint venture of quantum mechanics and general relativity
leads to a remarkable discovery that a "black hole (BH) behaves
like a black body". The thermal radiation emitted by a BH has
temperature proportional to its surface gravity at the horizon
while entropy is proportional to the area of the horizon $[1,2]$,
i.e., $T=\frac{\kappa}{2\pi}, ~~S=\frac{A}{4G}$. Also this
temperature entropy and the mass of the BH are related by the
first law of thermodynamics $[3]$. Further, the equivalence of the
thermodynamical quantities with the geometry of the horizon leads
to speculate some inherent relationship between thermodynamical
laws and Einstein equations. This speculation comes into true in
1995 when Jacobson $[4]$ was able to formulate Einstein's
equations from the Clausius relation ($\delta Q=TdS$) using local
Rindler Causal horizons with $T$, the Unruh temperature as seen by
an accelerated observer just inside the horizon. The equivalence
in other way round was shown by Padmanabhan $[5]$ for a general
spherical symmetric space time. However in other gravity theories,
such type of equivalence is not possible. Eling et al $[6]$ showed
that the usual Clausius relation and the entropy assumption
$S=\alpha f'(r) A$(or $S=\alpha F\left(\phi\right) A$) for
$f(R)$-gravity (or scalar tensor gravity) do not give the correct
equations of motion-on entropy production term has to be added to
the Clausius relation and as a result there will be
non-equilibrium thermodynamics of space-time $[6,7]$.\\

Similarly, in the context of BH thermodynamics most studies of BH
thermodynamics are concentrated to stationary BHs. It is
speculated that thermodynamics of dynamical (i.e., non stationary)
BH is related to the non-equilibrium thermodynamics of the
universe. Hayward $[8]$ initiated the study of the thermodynamics
of dynamical BH by proposing "Unified first law". He introduced
the idea of the trapping horizon and was able to show the
equivalence of Einstein equations and unified first law . He
formulated the first law of thermodynamics for a dynamical BH by
projecting unified first law along the trapping horizon. Also
projecting along the tangent to the trapping horizon he was able
to formulate the Clausius relation. Subsequently, Cai et al $[7]$
studied in details the thermodynamics of FRW universe starting
from the unified first law, in Einstein gravity, Lovelock gravity
and in scalar-tensor theory of gravity and showed the necessity of
introducing entropy production term in scalar tensor theory.\\

In this small article we study the unified first law in details
and reveal some of its properties in the form of lemmas.\\

Mathematically, the unified first law can be written as
\begin{equation}
dE=A\psi+WdV
\end{equation}
Here,
\begin{equation}
E=\frac{R}{2G}\left(1-h^{ab}\partial_{a}R\partial_{b}R\right),
\end{equation}
is known as Misner-sharp energy. this is the total energy inside
the sphere of radius $R$(known as area radius). The energy supply
$\psi$ and the work function $W$ are given by $[7,8]$
\begin{equation}
\Psi=\psi_{a}dx^{a}~~~~,~~~~W=-\frac{1}{2}Trace (T)
\end{equation}
with
\begin{equation}
\psi_{a}=T_{a}^{b}\partial_{b}R+\partial_{a}R
\end{equation}
known as energy-supply vector.
The four dimensional metric is written in the form
\begin{equation}
ds^{2}=h_{ab}dx^{a}dx^{b}+R^{2}d\Omega^{2}
\end{equation}
where $R$ is the radius of the 2-sphere, $h_{ab}$ is the metric on
the 2-D hyper surface normal to the spherical surface of symmetry.
A and V are the area and the volume of the 2-sphere of radius $R$.
In the above equations (3) and (4) $T_{a}^{b}$ is the projection
of the energy momentum tensor normal to the spherical surface and
trace is taken over normal 2D hyper surface.\\

Now projecting the unified first law of thermodynamics along the
tangent to the trapping horizon gives the first law of BH
thermodynamics as $[7,9]$.
\begin{equation}
\left<dE,~ z\right>=\frac{\kappa}{8\pi G}\left<dA,~ z\right>+W\left<dV,~ z\right>
\end{equation}

where $\kappa$ and z are respectively the surface gravity and
tangent vector to the trapping horizon. Also the Clausius relation
has the form,
\begin{equation}
\left<A \psi, ~z\right>=\frac{\kappa}{8\pi G}<dA, ~z>
\end{equation}
For simplicity, we consider homogeneous and isotropic FRW model of
the universe and the line element is written in the form of
equation (5). Then we have,
\begin{equation}
h_{ab}=diag\left(-1,~~\frac{a^{2}}{1-kr^{2}}\right)
\end{equation}
and $R=ar$, the area radius.\\\\
The explicit form of different thermodynamical parameters are
$$E=\frac{R^{3}}{2G}\left(H^{2}+\frac{k}{a^{2}}\right)$$
$$W=\frac{1}{2}\left(\rho-p\right)$$
\begin{equation}
\psi_{0}=-\frac{1}{2}\left(\rho+p\right)H R
\end{equation}
$$\psi_{1}=\frac{1}{2}\left(\rho+p\right)a$$
$$\kappa=\frac{1}{2\sqrt{-h}}\partial_{a}\left(\sqrt{-h}~h^{ab}\partial_{b}R\right)=\frac{2\pi R}{3}\left(3p-\rho\right)$$
where we have assumed perfect fluid as the matter in the universe.\\

The two null vectors defined along the normal to the spherical surface of symmetry are given by,
\begin{equation}
\xi_+=
\partial_{+}=-\sqrt{2}\left(\partial_{t}-\frac{\sqrt{1-kr^{2}}}{a}
\partial r\right)~~~~, ~~~~\xi_- =\partial_{-}=-\sqrt{2}\left
(\partial_{t}+\frac{\sqrt{1-kr^{2}}}{a}\partial r\right)
\end{equation}
and the trapping horizon is defined as
\begin{equation}
h^{ab}\partial_{a}R=0
\end{equation}
i.e.,$$R_{T}=\frac{1}{\sqrt{H^{2}+\frac{k}{a^{2}}}}=R_{A},$$
the apparent horizon.
Now we shall prove the following lemmas :\\

{\bf Lemma $I$ :  Trapping horizons (if exists) are always apparent horizons but not the converse.}\\

{\bf Proof  :}  The dynamical apparent horizon is the marginally
trapped surface with vanishing expansion and is defined as a
sphere of radius $R_{A}$ satisfying
\begin{equation}
h^{ab}\partial_{a}R\partial_{b}R=0
\end{equation}
A trapping horizon is a hyper surface foliated by marginal spheres and is mathematically defined as
\begin{equation}
\theta_{+}R=\frac{1}{R}\partial_{+}R=0
\end{equation}
i.e., in the form of the metric (5) we have
\begin{equation}
h^{ab}\partial_{a}R=0
\end{equation}
Here $\theta_+$ is the expansion of the null congruence
$\xi_+=constant$.\\

Hence
$$h^{ab}\partial_{a}R\partial_{b}R=\left(h^{ab}\partial_{a}R\right)\partial_{b}R=0$$
But $h^{ab}\partial_{a}R\partial_{b}R=0$ does not imply $h^{ab}\partial_{a}R=0$

Thus the trapping horizon is always implies an apparent horizon but not the converse.\\\\

{\bf Lemma $II$ :  Unified first law and both the Friedman equations are equivalent on any spherical surface of symmetry}\\

{\bf Proof  :}  From equation (9)
\begin{equation}
dE=\frac{1}{2G}\left[3\left(H^{2}+\frac{k}{a^{2}}\right)aR^{2}\left(Hdt+dr\right)+2HR^{3}\left(\dot{H}-\frac{k}{a^{2}}\right)dt\right]
\end{equation}
\begin{equation}
\Psi=\psi_{0}dt+\psi_{1}dr=\frac{1}{2}\left(\rho+p\right)\left[-HRdt+adr\right]
\end{equation}
So
$$A\Psi+WdV=2\pi R^{2}\left(\rho+p\right)\left[-HRdt+adr\right]+2\pi R^{2}\left(\rho-p\right)\left[HRdt+adr\right]$$
$$=4\pi R^{2}\left[-pHRdt+a\rho dr\right]$$
Hence equating coefficients of $dt$ and $dr$ in the unified first law (1) we have (after some simplification)
\begin{equation}
H^{2}+\frac{k}{a^{2}}=\frac{8\pi G}{3}\rho
\end{equation}
and
\begin{equation}
\dot{H}-\frac{k}{a^{2}}=-4\pi G \left(\rho+p\right),
\end{equation}
the two Friedman equations at any $R$.
Hence unified first law and Friedman equations are equivalent on any spherical surface.\\\\

{\bf Note : } In ref $[7]$ Cai et al stated that (see remarks
after eq. (4.14)) Unified first law is just an identity related to
the $(0,~0)$-th component of Einstein equations. But in the above
lemma we have shown that both the Friedman equations are derivable
from the unified first law and vice versa.\\\\

{\bf Lemma $III$ :  The validity of the Clausius relation depends on the choice of the tangent vector z.}\\

{\bf Proof  :} According to Hayward the Clausius relation has the form :
$$\left<A\Psi,~ z\right>=\frac{\kappa}{8\pi G}\left<dA, z\right>$$
i.e., $$2 \pi R^{2}\left(\rho+p\right)\left[-HRz_{1}+\sqrt{1-kr^{2}}z_{2}\right]=\frac{\kappa}{G}R\left[R H z_{1}+\sqrt{1-kr^{2}}z_{2}\right]$$
i.e.,
$$\frac{z_{1}}{z_{2}}=\frac{\sqrt{1-\frac{k}{a^{2}}R^{2}}}{R H}\left(\frac{2\rho}{\rho+3p}\right)$$
Thus if the coefficients $z_{1}, ~z_{2}$ of the tangent vector
satisfy the above relation then only Clausius relation holds,
i.e., at a definite spherical hyper surface only for a particular
tangent vector Clausius relation holds.\\\\

{\bf Lemma $IV$ :  The first law of BH thermodynamics is
equivalent to Clausius relation at any spherical surface.}\\

{\bf Proof  :}  The first law of BH thermodynamics in Hayward's formalism can be written as
\begin{equation}
\left<dE,~z\right>=\frac{\kappa}{8\pi G}\left<dA,~z\right>+W\left<dV,~z\right>
\end{equation}
Now writing explicitly both sides we have (we have after some simplifications)
$$\frac{z_{1}}{z_{2}}=\frac{\sqrt{1-\frac{K}{a^{2}}R^{2}}}{RH}.\frac{2\rho}{\left(\rho+3p\right)}$$
Thus on a particular spherical surface and for the same tangent
vector both first law of BH thermodynamics and Clausius relation
hold.\\\\

The present study shows that although the unified first law is
expressed in terms of thermodynamical parameters.,it is
essentially an alternative form of Einstein field equations. It is
found that projecting this unified first law along tangent to the
spherical surface of symmetry gives the equivalence of 1st law of
BH thermodynamics and Clausius relation but it holds only for a
specific choice of the tangent vector which may be different for
different spherical surfaces. For future work, it will be nice to
examine the equivalence of unified first law and the Einstein
field equations for general space-time model.\\

{\bf Acknowledgement :}\\

RB wants to thank West Bengal State Government for awarding JRF.
 NM wants to thank CSIR, India for awarding JRF. All the authors are
 thankful to IUCAA, Pune as this work was done during a visit.\\

{\bf REFERENCES}\\
\\
$[1]$ S.W.Hawking, \textit{Commun.Math.Phys} \textbf{43} 199
(1975).\\\\
$[2]$ J.D.Bekenstein, \it{Phys. Rev. D} {\bf 7} 2333 (1973).\\\\
$[3]$ J.M.Bardeen , B.Carter and S.W.Hawking, {\it
Commun.Math.Phys } {\bf 31} 161 (1973).\\\\
$[4]$ T.Jacobson, \it {Phys. Rev Lett.} {\bf 75} 1260 (1995). \\\\
$[5]$ T.Padmanabhan, \it {Class. Quantum Grav} {\bf 19} 5387
(2002); \it{Phys.Rept} {\bf 406} 49 (2005).\\\\
$[6]$ C.Eling, R.Guedens and T.Jacobson, \it {Phys. Rev Lett.}
{\bf 96} 121301 (2006). \\\\
$[7]$ R.G. Cai and L.M. Cao , {\it Phys. Rev. D} {\bf 75} 064008 (2007).\\\\
$[8]$ S.A. Haywards, \it{Phys. Rev. D} {\bf 49} 6467 (1994).\\\\
$[9]$ S.A. Haywards,  \it {Class. Quantum Grav} {\bf 15} 3147
(1998).\\

\end{document}